\documentclass[aps,prd,twocolumn,showpacs,showkeys,superscriptaddress]{revtex4-1}
\usepackage{latexsym}
\usepackage{amssymb}
\usepackage{amsmath}
\usepackage{amscd}
\usepackage{amsthm}
\usepackage{graphicx}
\usepackage{textcomp}
\usepackage{colortbl}
\usepackage{hyperref}
\usepackage[font={footnotesize,it}]{caption}
\usepackage{multirow}
\usepackage[T1]{fontenc}
\usepackage{ae,aecompl}
\begin{document}
\title{Current Constraints on Anisotropic and Isotropic Dark Energy Models} 

\author{Hassan Amirhashchi}
\email[]{h.amirhashchi@mhriau.ac.ir, hashchi@yahoo.com}
\affiliation{Department of Physics, Mahshahr Branch, Islamic Azad University,  Mahshahr, Iran}
\author{Soroush Amirhashchi}
\email[]{soroush.amirhashchi@gmail.com}
\affiliation{Department of Statistics, Shahid Beheshti University, Tehran, Iran}
\begin{abstract}
We use Gaussian processes in combination with MCMC method to place constraints on cosmological parameters of three dark energy models including flat and curved FRW and Bianchi type I spacetimes. To do so, we use recently compiled 36 measurements of the Hubble parameter $H(z)$ in the redshifts intermediate $0.07\leqslant z \leqslant 2.36$. Moreover, we use these models to estimate the redshift of the deceleration-acceleration transition. We consider two Gaussian priors for current value of the Hubble constant i.e $H_{0}=73\pm1.74 (68\pm 2.8)$ km/s/Mpc to investigate the effect of the assumed $H_{0}$ on our parameters estimations. For statistical analysis we use NUTS sampler which is an extension of Hamiltonian Monte Carlo algorithm to generate MCMC chains for parameters of dark energy models. To compare the considered cosmologies, we perform Akaike information criterion (AIC) and Bayes factor ($\Psi$). In general, when we compared our results with 9 years WMAP as well as Planck 2015 Collaboration, we found that Bianchi type I model is slightly fits better to the observational Hubble data with respect to the non-flat FRW model.
\end{abstract}
\smallskip
\maketitle
\section{Introduction}

The discovery of accelerated expanding universe by Riess et al \cite{ref1} and Perlmutter et al \cite{ref2} was a great shock to the scientific community. Although we knew that the universe is expanding, but due to the gravitational attractive force, this expansion was supposed to be speeding down. The accelerating expansion means that there should be an unknown energy component which works against gravity. In the context of General Theory of Relativity (GR), this unknown energy is called \textgravedbl dark energy \textacutedbl (DE). It is also believed that the behavior of universe could be investigated by modification of gravity and hence there is no need to consider an unknown component namely dark energy in the cosmic fluid. In the framework of GR, the study of DE is possible through it's equation of state parameter (EOS) which is defined as $\omega_{X}=\frac{p_{X}}{\rho_{X}}$, where $p_{X}$ and $\rho_{X}$ are the pressure and density of DE respectively. In general, the EOS parameter varies in three possible regions including: (1)  ($-\frac{1}{3}>\omega_{X}>-1$) which could be described by a scalar filed $\phi$ that is minimally coupled to gravity called quintessence \cite{ref3,ref4,ref5,ref6,ref7}, this scenario is not in good agreement with XCDM Model as recent observations \cite{ref8,ref9,ref10,ref11,ref12,ref13} indicate that $\omega_{X}<-1$ is allowed at $\%68$ confidence level, (2) ($\omega_{X}=-1$) which describes by cosmological constant $\Lambda$ \cite{ref14,ref15}. Although this scenario is the first and natural candidate for DE but it suffers from cosmological and coincidence problems \cite{ref16,ref17,ref18} (3) ($\omega_{X}<-1$) which could be described by a new class of scalar field models with negative kinetic energy called phantom \cite{ref19}. However, phantom fields are generally plagued by ultraviolet quantum instabilities \cite{ref20}. It is worth nothing that the possibility of $\omega_{X}<-2$ has been ruled out by the current observations \cite{ref11,ref21,ref22,ref23}. The existence of DE causes a cosmic expansion phase transition from decelerating to accelerating. This phase transition could be studied via the sign change of the universal deceleration parameter $q(z)$ which is restricted in interval $-1<q<0$. It is believed that the basic characteristics of the cosmological evolution could be expressed in terms of the Hubble parameter $H_{0}$ and the deceleration parameters $q_{0}$ \cite{ref24}. This in turn enable us to construct model-independent kinematics of the cosmological expansion. Based on what we discussed above, in our study, we derive the Hubble parameter $H(z)$ as a direct function of $H_{0}, q_{0}$ for all three models under study. This allow us to fit these parameters to data in a direct way not as transform parameters.\\

As far as we consider cosmological principle, we can work in the framework of FRW metric which is maximally symmetric (homogeneous and isotropic). From mathematical point of view, all three spatial components of this metric are the same functions of time (redshift). In fact this is the main and popular space-time where cosmologists use to study the behavior of our universe (specially DE) at the present epoch. Nevertheless, Recent observations indicate very tiny variations, one part in a hundred thousand, in the intensities of the microwaves coming from different directions in the sky. This observations challenge the isotropy assumption of FRW spacetime. Moreover, we are not sure that the spacetime of universe has the same properties as FRW metric beyond event horizon. A more realistic solution could be presented by considering models with different spatial components (at least these models provide an arena for testing the accuracy of FRW models). If these components are assumed to be only functions of time (not space), it means that we have ignored the isotropy of the metric. Such kind of homogeneous and anisotropic metrics are given by Bianchi Type models. The high idealization of FRW metric has been already motivated researchers to study non-FRW based universes such as Lema\^{i}tre-Tolman-Bondi (LTB) metric which is isotropic and inhomogeneous (for example see \cite{ref25,ref26,ref27,ref28,ref29,ref30}). Recently Farooq et al \cite{ref31} have made a detailed analysis of five FRW based cosmological models by using 38 Hubble parameter measurements. Soon after, Yu et al \cite{ref32} , because of the partial overlap of the WiggleZ and BOSS spatial regions (see Beutler et al \cite{ref33}), drop the three WiggleZ radial BAO points (Blake et al \cite{ref34}) from Table 1 of Farooq et al \cite{ref31} but include the recent redshift $z = 0.47$ cosmic chronometric measurement \cite{ref35}. In this paper, we use these 36$H(z)$ datapoints to make a detailed comparison between Bianchi type I dark energy model (XBI) and FRW (XCDM) universes. Very recently Magana et al \cite{ref36} have compiled a list of 51 observational Hubble data (OHD) in the redshift range $0.07\leqslant z \leqslant 2.36$ containing contains 31 data points measured with the differential age method by Jimenez and Loeb \cite{ref37}, and 20 data points obtained from clustering of galaxies. They used these data to study the Cardassian expansion model in which the late time cosmic acceleration is driven by the modification of the Friedmann equation as $H^{2}=f(\rho)$ \cite{ref38} where $\rho$ is the energy density of the Universe. The plane of this paper is as follows: In Section~\ref{sec:2} we give a brief discussion about three cosmological models under estimation. Section~\ref{sec:3} deals with the analytical method we used for the statistical analysis. In Subsecs~\ref{sec:3.1}, \ref{sec:3.2}, and \ref{sec:3.3} we constrain flat XCDM, curved XCDM, and anisotropic XBI dark energy models respectively. Derivations of the transition redshifts in three models are presented in In Section~\ref{sec:4}. We summarize our findings and conclude in In Section~\ref{sec:5}.
\section{Cosmological Models}\label{sec:2}
In this section we briefly describe the three models we use to analyze the 36$H(z)$ data. These are the non-flat XCDM model
that allows for spatial curvature and where dark energy represented by an X-fluid, as well as the
flat XCDM model which is the same as non-flat model except spatial curvature is zero in this case.
We also consider Bianchi type I dark energy, XBI, with dynamical dark energy. It is worth nothing that some Bianchi models isotropise due to inflation \cite{ref39}.\\
To derive first two models we can write Friedmann equations
\begin{equation}
	\label{eq1} \left(\frac{\dot{a}}{a}\right)^{2}=\frac{\rho_{m}+\rho_{X}}{3}-\frac{K}{a^{2}},
\end{equation}
\begin{equation}
	\label{eq2} 2\frac{\ddot{a}}{a}=-\frac{1}{3}(\rho_{m}+\rho_{X}+3p_{X}),
\end{equation}
in the following form
\begin{equation}
	\label{eq3} \rho_{X}=3H^{2}+3Ka^{-2}-\rho_{m},
\end{equation}
\begin{equation}
	\label{eq4} p_{X}=(2q-1)H^{2}-Ka^{-2},
\end{equation}
where we assumed $(8\pi G=c=1, \omega_{m}=0$), $H=\frac{\dot a}{a}$ is the Hubble parameter, $q=-\frac{\ddot a}{aH^{2}}$ is the deceleration parameter, K is the spatial curvature, $(\rho_{m},\rho_{X})$ are the energy densities of cold dark matter (DM) and DE respectively. Using eqs (\ref{eq3},\ref{eq4}) we can obtain dark energy EOS parameter as
\begin{equation}
\label{eq5} \omega_{X}=\frac{1}{3}\frac{2q-1+\Omega_{0K}(1+z)^{2}\left(\frac{H}{H_{0}}\right)^{-2}}{1-\Omega_{0K}(1+z)^{2}\left(\frac{H}{H_{0}}\right)^{-2}-\Omega_{0m}(1+z)^{3}\left(\frac{H}{H_{0}}\right)^{-2}},
\end{equation}
where $a=(1+z)^{-1}$. Considering
\[
\Omega_{0K}=-\frac{K}{H_{0}^{2}},~~~~~\Omega_{0m}=\frac{\rho_{0m}}{3H_{0}^{2}},
\]
from eq(\ref{eq5}) we can find the Hubble parameter 
\begin{equation}
\label{eq6} H(z; {\bf P})=H_{0}\left[\frac{(1+3\omega_{X})\Omega_{0K}(1+z)^{2}+3\Omega_{0m}\omega_{X}(1+z)^{3}}{1+3\omega_{X}-2q}\right]^{\frac{1}{2}},
\end{equation}
where ${\bf P}$ is the set of the model parameters to be estimated from fitting of 36$H(z)$ datapoints. Above equation shows the general form of XCDM model (non-flat), by putting $\Omega_{k}=0$ we obtain flat XCDM model as
\begin{equation}
	\label{eq7} \omega_{X}=\frac{1}{3}\frac{2q-1}{1-\Omega_{0m}(1+z)^{3}\left(\frac{H}{H_{0}}\right)^{-2}},
\end{equation}
\begin{equation}
	\label{eq8} H(z; {\bf P})=H_{0}\left[\frac{3\Omega_{0m}\omega_{X}(1+z)^{3}}{1+3\omega_{X}-2q}\right]^{\frac{1}{2}}.
\end{equation}
\\

Also the \textgravedbl Friedmann-like\textacutedbl equations for Bianchi type I model could be written as (\cite{ref40,ref41} )
\begin{equation}
\label{eq9} \left(\frac{\dot{a}}{a}\right)^{2}=\frac{\rho_{m}+\rho_{X}}{3}+\frac{\tilde{K}}{a^{6}},
\end{equation}
\begin{equation}
\label{eq10} 2\frac{\ddot{a}}{a}=-\frac{1}{3}(\rho_{m}+\rho_{X}+3p_{X}),
\end{equation}
where $\tilde{K}$ indicates the anisotropy of the model (see \cite{ref42} for details). Note that the anisotropy parameter $\tilde{K}$ does not appear in (\ref{eq10}) as it canceled out in the derivation. From above two equations we obtain
\begin{equation}
\label{eq11} \rho_{X}=3H^{2}-3\tilde{K}a^{-6}-\rho_{m},
\end{equation}
\begin{equation}
\label{eq12} p_{X}=(2q-1)H^{2}-\tilde{K}a^{-6}.
\end{equation}
Hence, in this case, the EOS parameter of DE is given by
\begin{equation}
\label{eq13} \omega_{X}=\frac{1}{3}\frac{2q-1+\Omega_{0\tilde{K}}(1+z)^{6}\left(\frac{H}{H_{0}}\right)^{-2}}{1-\Omega_{0\tilde{K}}(1+z)^{6}\left(\frac{H}{H_{0}}\right)^{-2}-\Omega_{0m}(1+z)^{3}\left(\frac{H}{H_{0}}\right)^{-2}}.
\end{equation}
From this equation we obtain the Hubble parameter as
\begin{equation}
\label{eq14} H(z; {\bf P})=H_{0}\left[\frac{(1+3\omega_{X})\Omega_{0\tilde{K}}(1+z)^{6}+3\Omega_{0m}\omega_{X}(1+z)^{3}}{1+3\omega_{X}-2q}\right]^{\frac{1}{2}}.
\end{equation}
It is interesting to note that for all three models (from eqs \ref{eq5}, \ref{eq7}, \ref{eq13}), the ultimate fate of our universe describe by following EOS parameter
\begin{equation}
\label{eq15} \omega_{X}=\frac{2q-1}{3},~~~~\mbox{at}~~~~ z=-1.
\end{equation}
Obviously eq \ref{eq15} could be considered as a fundamental equation which describes the state of universe at its final stages (the same equation has been recently found for Bianchi type V model \cite{ref30}. It is also worth to mention that as $q\to -1$, $\omega_{X}\to -1$ which is corresponding to cosmological constant scenario. In other word, all the models ultimately settle into a de-Sitter phase. We also use the following equation to compute the age (in Gyr) of universe given by each of models.
\begin{equation}
\label{eq16} t=\frac{10^3}{H_{0}}\int_{0}^{\infty}\dfrac{dz}{(1+z)\sqrt{\frac{(1+3\omega_{X})\tilde{\Omega}(1+z)^{n}+3\Omega_{0m}\omega_{X}(1+z)^{3}}{1+3\omega_{X}-2q}}},
\end{equation}
where the pair $(\tilde{\Omega}, n)$ is $(0, 2)$, $(\Omega_{0{K}}, 2)$, and $(\Omega_{0\tilde{K}}, 6)$ for flat XCDM, curved XCDM, and XBI models respectively.\\

In Sec~\ref{sec:3} we use these expressions for the Hubble parameter in conjunction with observational Hubble data (OHD) to constrain the cosmological parameters of above mentioned dark energy models.
\section{Method and Results}\label{sec:3}
We use the recently compiled $36H(z)$ datapoints \cite{ref32} in the redshift range $0.07\leq z\leq 2.36$ to constrain our model parameters (see Table~\ref{tab:1}). As mentioned before, we have dropped three correlated date from Table 1 of Farooq et al \cite{ref31} but include the recent redshift $z = 0.47$ cosmic chronometric measurement by Ratsimbazafy et al \cite{ref35}.  
\begin{table}
\centering
\caption{Hubble parameter versus redshift data.}
\setlength{\tabcolsep}{30pt}
\scalebox{0.6}{
\begin{tabular} {cccc}
\hline
\hline
$H(z)$    &  $\sigma_{H}$   &  $z$  & Reference\\[0.5ex]
			
\hline{\smallskip}
89    &   50       &0.47     & \cite{ref35} \\
			
69 &      19.6     & 0.070   & \cite{ref55}\\
			
69 &      12       & 0.090   & \cite{ref56} \\
			
68.6 &    26.2     & 0.120   & \cite{ref55}\\
			
83 &      8        & 0.170   & \cite{ref56} \\
			
75 &      4        & 0.179   & \cite{ref57} \\
			
75 &      5        & 0.199   & \cite{ref57} \\
			
72.9 &    29.6     & 0.200   & \cite{ref55}\\
			
77 &      14       & 0.270   & \cite{ref56} \\
			
88.8 &    36.6     & 0.280   & \cite{ref55}\\ 
			
83  &     14       &0.352    & \cite{ref57} \\
			
81.5  &   1.9      &0.380    & \cite{ref49} \\
			
83  &     13.5     &0.3802   & \cite{ref59} \\
			
95  &     17       &0.400    & \cite{ref56} \\
			
77  &     10.2     &0.4004   & \cite{ref59} \\
			
87.1  &   11.2     &0.4247   & \cite{ref59} \\
			
92.8  &   12.9     &0.4497   & \cite{ref59} \\
			
80.9  &   9        &0.4783   & \cite{ref59} \\
			
97 &      62       & 0.480   &\cite{ref60}  \\
			
90.4  &   1.9       &0.510   &\cite{ref59} \\
			
104 &     13       &0.593    & \cite{ref57} \\
			
97.3  &   2.1      &0.610    & \cite{ref49} \\
			
92 &      8        &0.680    &\cite{ref57}  \\
			
105 &     12       &0.781    & \cite{ref57} \\
			
125 &     17       &0.785    & \cite{ref57} \\
			
90 &      40       &0.880    & \cite{ref60} \\
			
117 &     23       &0.900    & \cite{ref56} \\ 
			
154 &     20       &1.037    & \cite{ref57} \\
			
168 &     17       & 1.300   & \cite{ref56} \\ 
			
160  &    33.6     &1.363    & \cite{ref58} \\
			
177 &     18       &1.430    & \cite{ref56} \\
			
140 &     14       &1.530    & \cite{ref56} \\
			
202 &     40       &1.750    & \cite{ref56} \\
			
186.5 &   50.4     &1.965    & \cite{ref58} \\
			
222  &    7        &2.340    & \cite{ref61} \\ 
			
226  &    8       &2.360     & \cite{ref62}\\
			
\hline
\hline
\end{tabular}}
\label{tab:1}
\end{table}
Using Gaussian Process method (GP) one could determine continuous function of H(z) that best represents the discrete Hubble
parameter data  of Table~\ref{tab:1}. This continuous H(z) with baryon acoustic oscillation distance-redshift observations will be used to constrain parameters of DE models. The application of GP method in the study of DE was first used by Holsclaw et al \cite{ref43,ref44,ref45}, Shafieloo et al \cite{ref46}, and Seikel et al \cite{ref47,ref48}. Yu et al \cite{ref32} have recently used GP method to constrain some parameters of $\Lambda$CDM model. In general, while the Gaussian distribution is the distribution of a random variable, GP describes a distribution over functions. Assuming the
errors ($\sigma_{i}$) are Gaussian, we can describe the observational data $(z_{i},y_{i})$ by a Gaussian process in which the observed data are considered to be scattered around the underlying function $y_{i}=f(z_{i})+\sigma_{i}$. It is clear that the evaluated value of $y=f(z)$ at a point $z$ is a Gaussian random variable with mean $\mu(z)$ and variance $\sigma(z)$ given by
\begin{equation}
\label{eq17} \mu(z)=\sum_{i,j}^{N}k(z,z_{i})(M^{-1})_{ij}f(z_{j})~~~~ i,j=1,2,\dots N,
\end{equation}	
and
\begin{equation}
\label{eq18} \sigma(z)=k(z,z)-\sum_{i,j}^{N}k(z,z_{i})(M^{-1})_{ij}k(z_{j},z),
\end{equation}	
where $k(z,z_{i})=Cov(f(z_{i}),f(z_{j}); i\neq j$, is the covariance function which could be define in a Gaussian form as (for other chooses see \cite{ref32})
\begin{equation}
\label{eq19} k(z,\acute{z})=\sigma^{2}_{f}\exp\left[\frac{(z-\acute{z})^{2}}{2\ell^{2}}\right],
\end{equation}	
where length scale $\ell$ indicates distance between two $z$ points with significant $\Delta f(z)$ (i.e it is a measure of the coherence length of the correlation) whereas $\sigma_{f}$ defines the overall amplitude of the correlation. In above equations $M_{ij}=k(z_{i},z_{j})+c_{ij}$, where $c_{ij}$ is the covariance matrix of the observed data. Since three galaxy distribution radial BAO H(z) measurements \cite{ref49} are correlated, from Table~\ref{tab:1}, this correlation Matrix is given by 
\begin{equation}
\label{eq20} c=
\begin{bmatrix}
3.65 & 1.78 & 0.93  \\
1.78 & 3.65 & 2.20\\
0.93 & 2.20 & 4.45 
\end{bmatrix}.
\end{equation}
It is worth nothing that for uncorrelated data we have $c_{ij}=\mbox{diag}(\sigma_{i}^{2})$.\\
	
On the bases of Bayesian estimation methods, we are able to estimate the hyperparameters ($\sigma_{f},\ell$) of the GP correlation function together with any other parameters of a given model. To do so, we minimizing the following log marginal likelihood function \cite{ref47}
\[
\ln\mathcal{L}=-\frac{1}{2}\sum_{i,j}^{N}\left[f(z_{i})-\mu(z_{i})\right](M^{-1})_{ij}\left[f(z_{j})-\mu(z_{j})\right]
\]
\begin{equation}
\label{eq21} -\frac{1}{2}\ln[(2\pi)^{N}|M|]\\
\end{equation}
where $|M|$ is the determinant of matrix $M_{ij}$. We use NUTS sampler which is an extension of Hamiltonian Monte Carlo (HMC) algorithm to generate MCMC chains for all parameters of three models. We consider two Gaussian priors for current value of Hubble constant, $H_{0}=68\pm 2.8$ km/s/Mpc \cite{ref50} and $H_{0}=73\pm1.74 $ km/s/Mpc \cite{ref51} to study the effect of the assumed $H_{0}$ value on our results. Moreover, we assume a normal distribution for the current value of EOS parameter such as $\omega_{X}=-1\pm0.3$ but for all other parameters we assume the following uniform priors:
\[
\Omega_{m}\sim U(0,2), ~~~\Omega_{k} \sim U(-0.5, 0.5),~~~\Omega_{\tilde{k}} \sim U(-0.5, 0.5),
\]
	
\[ 
q \sim U(-1,0),~~~\sigma_{f} \sim U(0,100),~~~ \ell\sim U(0,10).
\]
To analyze MCMC chains, we use GetDist python package.
we confirm that the MCMC chains	converged by monitoring the trace plots and checking for good mixing and stationarity of the posterior distributions. In all our computations we consider four chains each of them with 25000 iterations for each parameter to stabilize the estimations. In next section, we compare our results to the Planck 2015 collaboration \cite{ref8} and 9WMAP \cite{ref10} (Table~\ref{tab:2}) to find out which DE model is in better agreement with these observations.
\begin{table}[ht]
\caption{Results from 9years WMAP and Planck 2015 collaboration for $\Lambda$CDM model at 1$\sigma$ confidence level.}
\centering
\setlength{\tabcolsep}{2pt}
\scalebox{0.7}{
\begin{tabular} {ccc}
\hline
Parameter    & WMAP+eCMB+BAO+H$_{0}$& TT+TE+EE+lensing+BAO+JLA+H$_{0}$ \\[0.5ex]
\hline
\hline{\smallskip}
$H_{0}$ &  $68.92^{+0.94}_{-0.95}$ &    $67.74\pm0.46$ \\[0.2cm] 
				
$\Omega_{m}$ &  $0.2855^{+0.0096}_{-0.0097}$ &  $0.3089\pm0.0062$ \\[0.2cm] 
				
$\Omega_{X}$ & $0.717\pm0.011$ & $0.6911\pm0.0062$\\[0.2cm] 
				
$\Omega_{k}$ & $-0.0027^{+0.0039}_{-0.0038}$ &$0.0008^{+0.0040}_{-0.0039}$\\[0.2cm] 
				
$\omega_{X}$ & $-1.073^{+0.090}_{-0.089}$ & $-1.019^{+0.075}_{-0.080}$\\[0.2cm] 			
				
$t_{0}$ & $13.88\pm0.16$& $13.799\pm0.021$ \\[0.5ex]
\hline
\hline
\end{tabular}}
\label{tab:2}
\end{table}
\subsection{Constraints on Flat XCDM Model}\label{sec:3.1}
According to eq \ref{eq8} we could directly estimate the following four parameters for flat XCDM model from 36 Hubble parameter measurements. It is worth to mention that, in general, the deceleration parameter is not a free parameter and it could be calculated based on the computed values of $\Omega_{m}$ and $\omega_{X}$. Hence, in this model only $H_{0}, \Omega_{m}$ and $\omega_{X}$ are independent parameters. 
\begin{equation}
\label{eq22} {\bf P}=\{\Omega_{m},\omega_{X},H_{0},q\}.
\end{equation}
		
\begin{figure}[h!]
\centering
\includegraphics[width=9cm,height=9cm,angle=0]{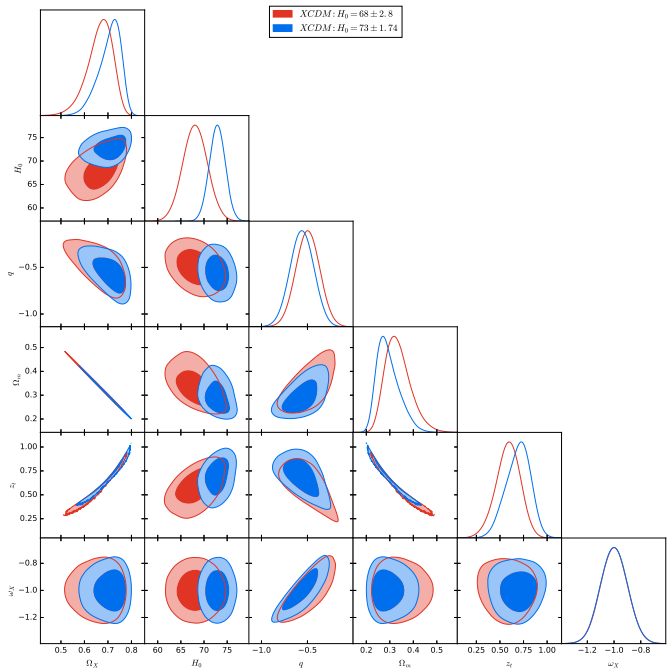}
\caption{One-dimensional marginalized distribution, and two-dimensional contours with $68\%$ CL and $95\%$ CL for flat XCDM model considering two normal $H_{0}$ priors.}
\label{fig1}
\end{figure}

\begin{figure}[h!]
\includegraphics[width=8cm,height=6cm,angle=0]{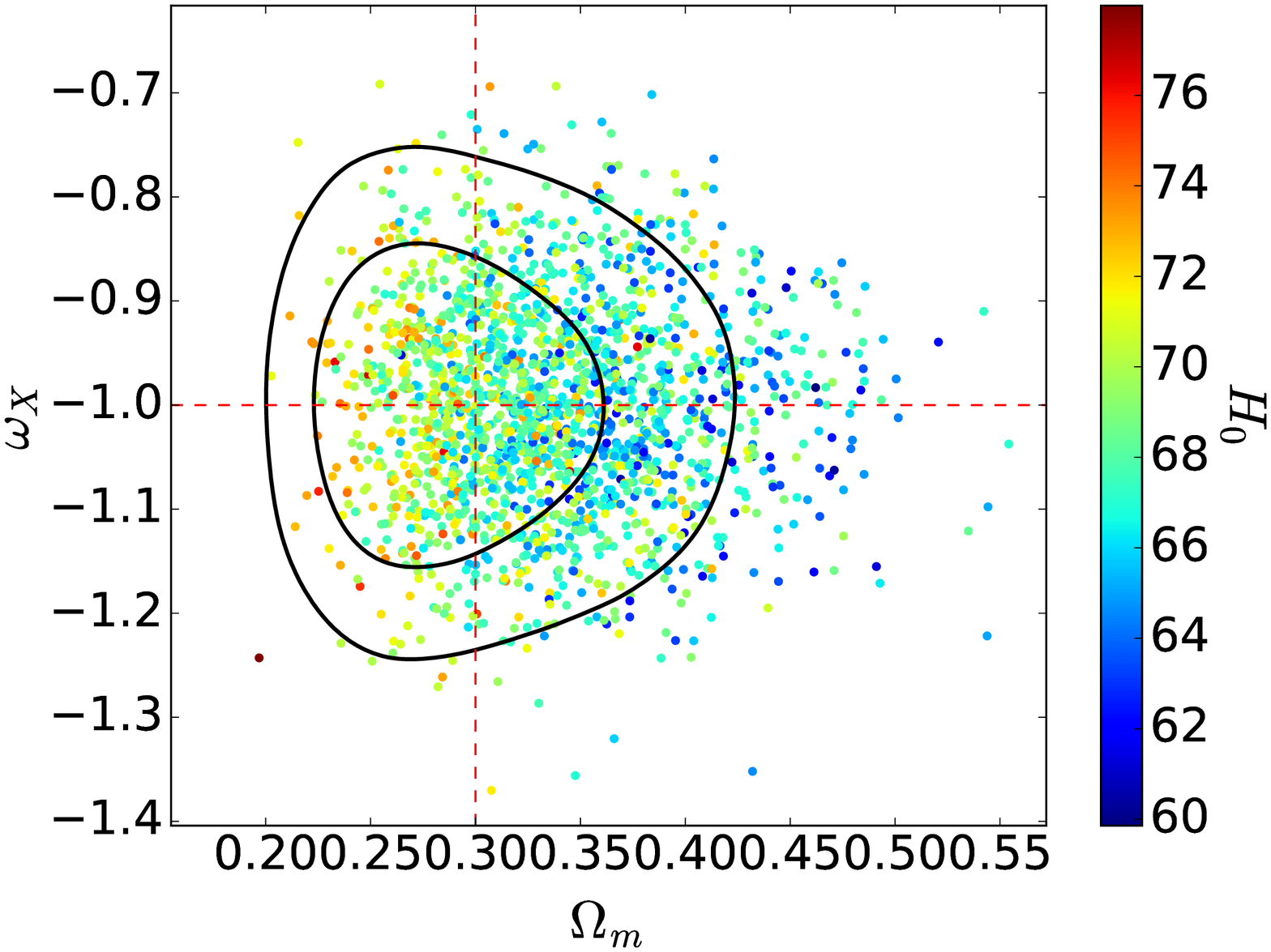} \\
\caption{Constraints in the $\omega_{X}-\Omega_{m}$ plane with $1\sigma$ and $2\sigma$ confident level in the flat XCDM model considering two normal $H_{0}$ priors. Samples are colored by the parameter $H_{0}$. Solid contours stand for prior $H_{0}=73\pm1.74$. Red-dashed horizontal \& vertical lines stand for $\omega_{X}=-1$ \& $\Omega_{m}=0.3$.}
\label{fig2}
\end{figure}

\begin{figure}[h!]
\includegraphics[width=8cm,height=6cm,angle=0]{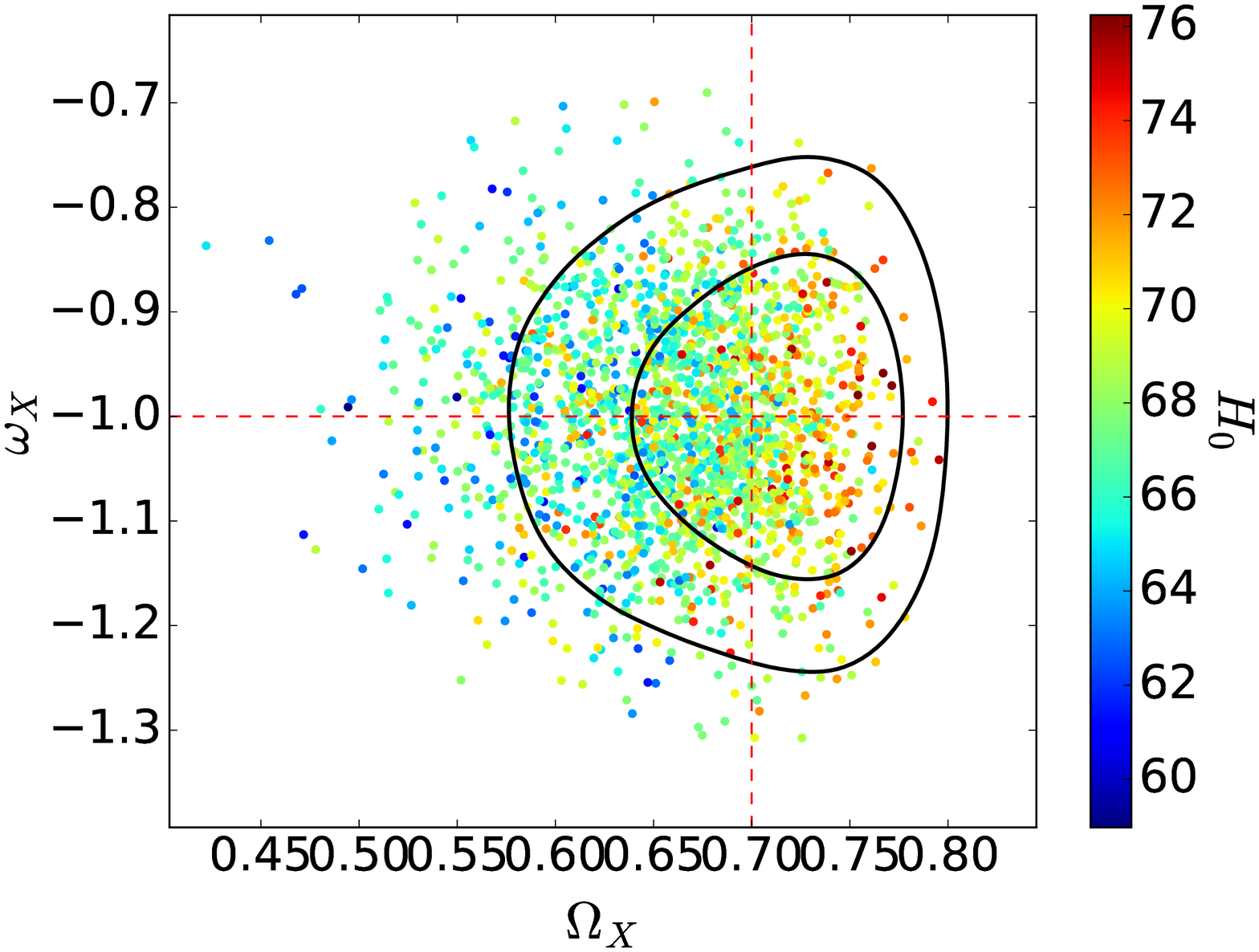}
\caption{Constraints in the $\omega_{X}-\Omega_{X}$ plane with $1\sigma$ and $2\sigma$ confident level in the flat XCDM model considering two normal $H_{0}$ priors. Samples are colored by the parameter $H_{0}$. Solid contours stand for prior $H_{0}=73\pm1.74$. Red-dashed horizontal \& vertical lines stand for $\omega_{X}=-1$ \& $\Omega_{X}=0.7$.}
\label{fig3}
\end{figure}

The results of our statistical analysis for this model are presented in Table~\ref{tab:3}. Also, Fig.~\ref{fig1} depicts the contour plots of the model parameters. Figs.~\ref{fig2} \& \ref{fig3} show the $1\sigma$ and $2\sigma$ contour plot of $(\omega_{X},\Omega_{m})$ \& $(\omega_{X},\Omega_{X})$ pairs respectively. From Table~\ref{tab:3} we observe that although our results for flat XCDM model are in good agreement with both the 9WMAP \cite{ref10} and Planck 2015 collaboration \cite{ref8}, but assuming $H_{0}=73\pm1.74$ places better constraints on the model parameters. Depending on the value of  $H_{0}=68\pm2.8 (73\pm1.74)$, the estimated value of deceleration parameter is obtained as $q=-0.50\pm 0.13 (-0.56\pm 0.13) $. In general, taking $H_{0}=73\pm1.74$ as prior for current Hubble expansion rate, the flat XCDM model behaves most likely as flat $\Lambda$CDM model. 

\begin{table}
\caption{Results from the fits of the flat XCDM model to the data at 1$\sigma$, 2$\sigma$, 3$\sigma$ confidence levels.}
\centering
\setlength{\tabcolsep}{2pt}
\scalebox{0.7}{
\begin{tabular}{c p{1cm} cccc}
\hline
\hline
& & Parameter & $\%68$ & $\%95$ & $\%99$ \\
\hline
\multirow{14}{*}{$H_{0}=68\pm 2.8$} &
\multirow{1}{*}{Fit} &
			
$H_{0}$ &  $68.11\pm 2.6$ & $68.1^{+5.2}_{-5.2}$ &  $68.1^{+6.9}_{-6.7}$ \\[.2cm]
&  & $\Omega_{m}$ &  $0.334^{+0.036}_{-0.071}$ & $0.335^{+0.11}_{-0.098}$ &  $0.34^{+0.17}_{-0.11}$ \\[.2cm]
& & $\omega_{X}$ & $-1.00\pm 0.09 $ & $-1.00^{+0.20}_{-0.20}$ & $-1.00^{+0.26}_{-0.26}$ \\[.2cm]
\cline{2-6}
& \multirow{2}{*}{Derived} &
$q$ & $-0.51\pm 0.11$  &  -  & - \\[.2cm] 
&& $t_{0}$ & $9.78\pm 2.41$ & - & -  \\[.2cm]
& & $\Omega_{X}$ & $0.665^{+0.063}_{-0.040}$ & - & -  \\[.2cm]
\hline
\hline
\multirow{14}{*}{$H_{0}=73\pm 1.74$} &
\multirow{1}{*}{Fit} &
$H_{0}$ &  $73.1\pm 1.4$ &    $72.9^{+3.4}_{-3.4}$ &  $72.9^{+4.4}_{-4.4}$ \\[.2cm]
& &$\Omega_{m}$ &  $0.292^{+0.033}_{-0.054}$   & $0.294^{+0.099}_{-0.082}$ &  $0.294^{+0.15}_{-0.091}$ \\[.2cm]
& &$\omega_{X}$ & $-1.00\pm 0.09$ & $-1.00^{+0.20}_{-0.19}$ & $-1.00^{+0.26}_{-0.26}$ \\[.2cm]
\cline{2-6}
& \multirow{2}{*}{Derived} &
$q$ & $-0.55\pm 0.63$ & - & - \\[.2cm]
&&$t_{0}$ & $9.12\pm 1.65$  &  -  & - \\[.2cm] 
& &$\Omega_{X}$ & $0.706^{+0.058}_{-0.032}$ & - & - \\[.2cm]
\hline
\hline
\end{tabular}}
\label{tab:3}
\end{table}
\subsection{Constraints on Non-Flat XCDM Model}\label{sec:3.2}
\begin{figure}[h!]
\centering
\includegraphics[width=9cm,height=9cm,angle=0]{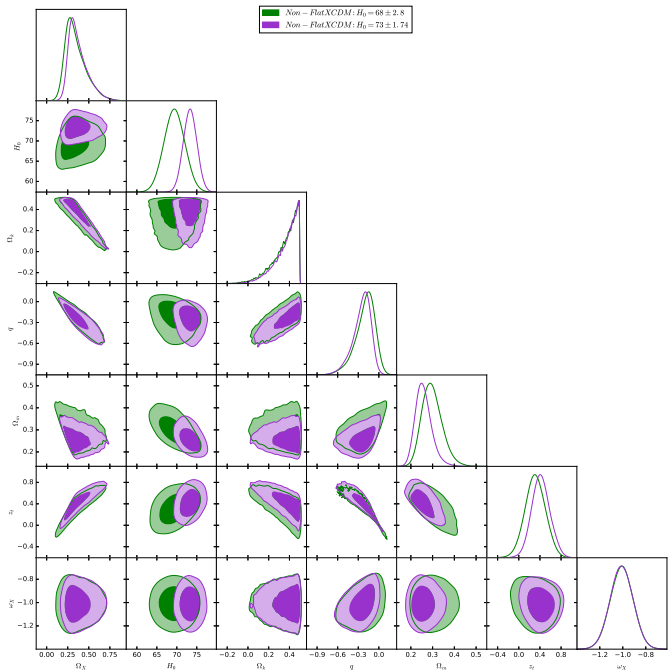}
\caption{One-dimensional marginalized distribution, and two-dimensional contours with $68\%$ CL and $95\%$ CL for flat curved XCDM model considering two normal $H_{0}$ priors.}
\label{fig4}
\end{figure}

\begin{figure}[h!]
\includegraphics[width=8cm,height=6cm,angle=0]{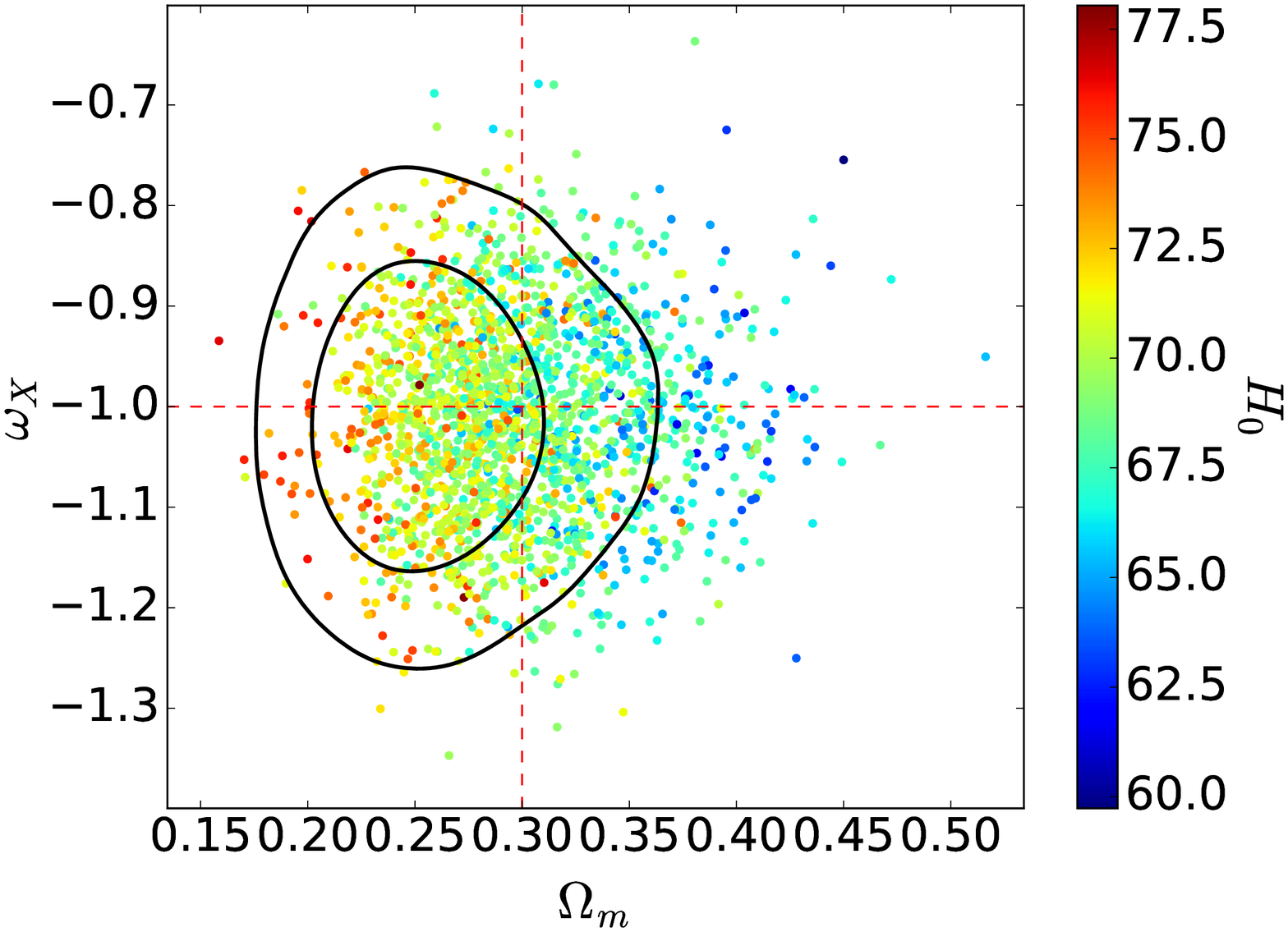} \\
\caption{$\omega_{X}-\Omega_{m}$ plane with $1\sigma$ and $2\sigma$ confident level in the curved XCDM model considering two normal $H_{0}$ priors. Samples are colored by the parameter $H_{0}$. Solid contours stand for prior $H_{0}=73\pm1.74$. Red-dashed horizontal \& vertical lines stand for $\omega_{X}=-1$ \& $\Omega_{m}=0.3$.}
\label{fig5}
\end{figure}

The non-flat XCDM model has five unknown parameter to be estimated from 36 Hubble parameter measurements. The base parameters set for this model is
\begin{equation}
\label{eq23} {\bf P}=\{\Omega_{m},\Omega_{0K},\omega_{X},H_{0},q\}.
\end{equation}
Table~\ref{tab:4} shows our results of statistical analysis for curved XCDM model. The contour plots of the model parameters obtained from the fit of the model to the OHD dataset at $1\sigma$ and $2\sigma$ confidence regions have been depicted in Fig.~\ref{fig1}.  Comparing the results of Table~\ref{tab:4} with Table~\ref{tab:2}, shows that the computed values of $q$, $\Omega_{X}$, and $\Omega_{K}$ are not in good agreement with 9 years WMAP as well as Planck 2016 collaboration. Specially, the estimated value of curvature density, $\Omega_{k}$, is very higher than what is expected. LHuillier and Shafieloo \cite{ref52} and Mitra et al\cite{ref53} have recently shown that the currently available non-CMB measurements do not significantly constrain spatial curvature. Moreover, As mentioned in ref \cite{ref54}, unlike the CMB anisotropy data, the H(z) data are not sensitive to the behavior of cosmological spatial inhomogeneities. Hence, more precise measurements of $H(z)$ at higher redshift are needed for tighter constraints on $\Omega_{K}$. For a closer view we have depicted the contour plots of $\omega_{X}-\Omega_{m}$ \& $\omega_{X}-\Omega_{X}$ pairs at Figs.~\ref{fig4} \& \ref{fig5}. Depending on the special chose of the current value of Hubble constant i.e $H_{0}=68\pm2.8 (73\pm1.74)$ the present value of the deceleration parameter is estimated as $q=-0.20^{+0.17}_{-0.10} (-0.252^{+0.16}_{-0.089})$. In general, although, same as flat XCDM case, choosing $H_{0}=73\pm1.74$ as prior improved our estimations but using only OHD dataset, generally, dose not enough to constrain this model. In fact, in addition to higher redshift data, the joint combination of different datasetes gives rise to better estimations.
\begin{figure}[h!]
\includegraphics[width=8cm,height=6cm,angle=0]{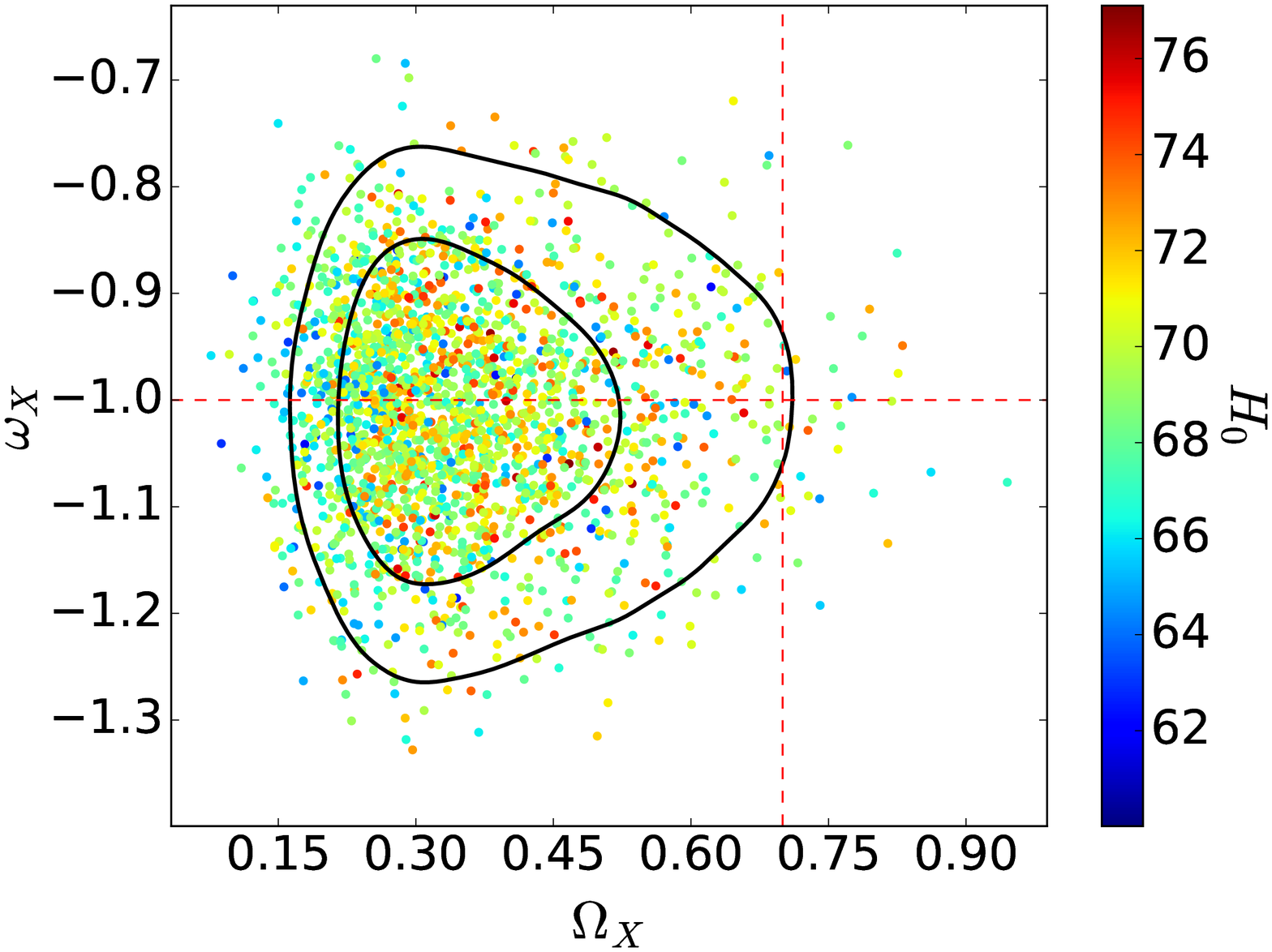}
\caption{$\omega_{X}-\Omega_{m}$ plane with $1\sigma$ and $2\sigma$ confident level in the curved XCDM model considering two normal $H_{0}$ priors. Samples are colored by the parameter $H_{0}$. Solid contours stand for prior $H_{0}=73\pm1.74$. Red-dashed horizontal \& vertical lines stand for $\omega_{X}=-1$ \& $\Omega_{X}=0.7$.}
\label{fig6}
\end{figure}

\begin{table}[h!]
\caption{Results from the fits of the curved XCDM model to the data at 1$\sigma$, 2$\sigma$, 3$\sigma$ confidence levels.}
\centering
\setlength{\tabcolsep}{2pt}
\scalebox{0.7}{
\begin{tabular}{c p{1cm} cccc}
\hline
\hline
& & Parameter & $\%68$ & $\%95$ & $\%99$ \\
\hline
\multirow{16}{*}{$H_{0}=68\pm 2.8$} &
\multirow{1}{*}{Fit} &
$H_{0}$ &  $69.4\pm 2.6 $ & $69.4^{+5.1}_{-5.0}$ &  $69.4^{+6.7}_{-6.6}$ \\[.2cm]
&  & $\Omega_{m}$ &  $0.298^{+0.038}_{-0.052}$ & $0.298^{+0.094}_{-0.087}$ &  $0.30^{+0.14}_{-0.10}$ \\[.2cm]
& & $q$ & $-0.20^{+0.17}_{-0.10}$  &  $-0.20^{+0.26}_{-0.31} $  & $-0.20^{+0.30}_{-0.48}$ \\[.2cm] 
& & $\omega_{X}$ & $-1.009\pm 0.099 $ & $-1.01^{+0.19}_{-0.20}$ & $-1.01^{+0.26}_{-0.26}$ \\[.2cm]
& & $\Omega_{k}$ & $0.356^{+0.14}_{-0.038} $ & $0.36^{+0.15}_{-0.25}$ & $0.36^{+0.15}_{-0.41}$ \\[.2cm]
\cline{2-6}
& \multirow{2}{*}{Derived} &
$t_{0}$ & $13.96\pm 2.13$ & - & -  \\[.2cm]
& & $\Omega_{X}$ & $0.346^{+0.074}_{-0.15}$ & - & -  \\[.2cm]
\hline
\hline
\multirow{16}{*}{$H_{0}=73\pm 1.74$} &
\multirow{1}{*}{Fit} &
$H_{0}$ &  $73.3\pm 1.7$ &    $73.3^{+3.4}_{-3.4}$ &  $73.3^{+4.4}_{-4.4}$ \\[.2cm]
& &$\Omega_{m}$ &  $0.259^{+0.029}_{-0.042}$   & $0.259^{+0.077}_{-0.071}$ &  $0.259^{+0.13}_{-0.081}$ \\[.2cm]
& &$q$ & $-0.252^{+0.16}_{-0.089}$ & $-0.25^{+0.23}_{-0.29}$ & $-0.25^{+0.27}_{-0.45}$ \\[.2cm]
& &$\omega_{X}$ & $-1.011\pm 0.099 $ & $-1.01^{+0.19}_{-0.19}$ & $-1.01^{+0.26}_{-0.26}$ \\[.2cm]
& & $\Omega_{k}$ & $0.366^{+0.13}_{-0.045}$ & $0.37^{+0.14}_{-0.23}$ & $0.37^{+0.14}_{-0.37}$ \\[.2cm]
\cline{2-6}
& \multirow{2}{*}{Derived} &
$t_{0}$ & $12.78\pm 2.12$  &  -  & - \\[.2cm] 
& &$\Omega_{X}$ & $0.376^{+0.063}_{-0.14}$ & - & - \\[.2cm]
\hline
\hline
\end{tabular}}
\label{tab:4}
\end{table}
\subsection{Constraints on XBI Model}\label{sec:3.3}
\begin{figure}[h!]
\centering
\includegraphics[width=9cm,height=9cm,angle=0]{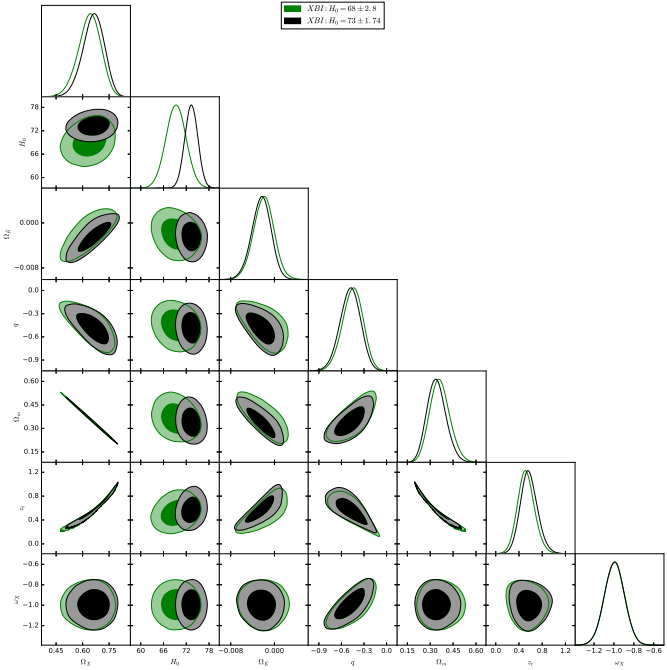}
\caption{One-dimensional marginalized distribution, and two-dimensional contours with $68\%$ CL and $95\%$ CL for XBI model considering two normal $H_{0}$ priors.}
\label{fig7}
\end{figure}

\begin{figure}[h!]
\includegraphics[width=8cm,height=6cm,angle=0]{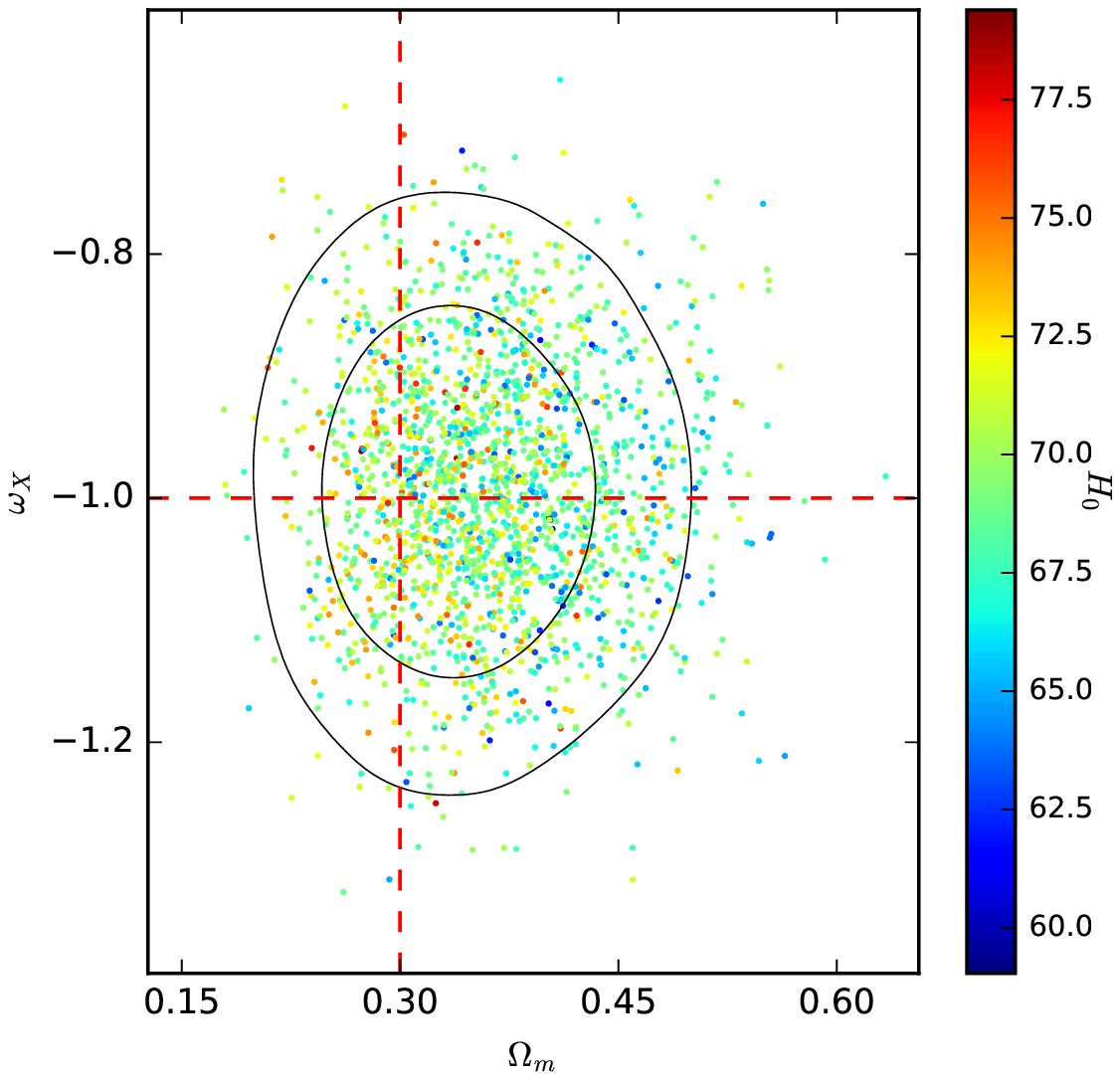} \\
\caption{$\omega_{X}-\Omega_{m}$ plane with $1\sigma$ and $2\sigma$ confident level in the XBI model considering two normal $H_{0}$ priors. Samples are colored by the parameter $H_{0}$. Solid contours stand for prior $H_{0}=73\pm1.74$. Red-dashed horizontal \& vertical lines stand for $\omega_{X}=-1$ \& $\Omega_{m}=0.3$.}
\label{fig8}
\end{figure}

\begin{figure}[h!]
\includegraphics[width=8cm,height=6cm,angle=0]{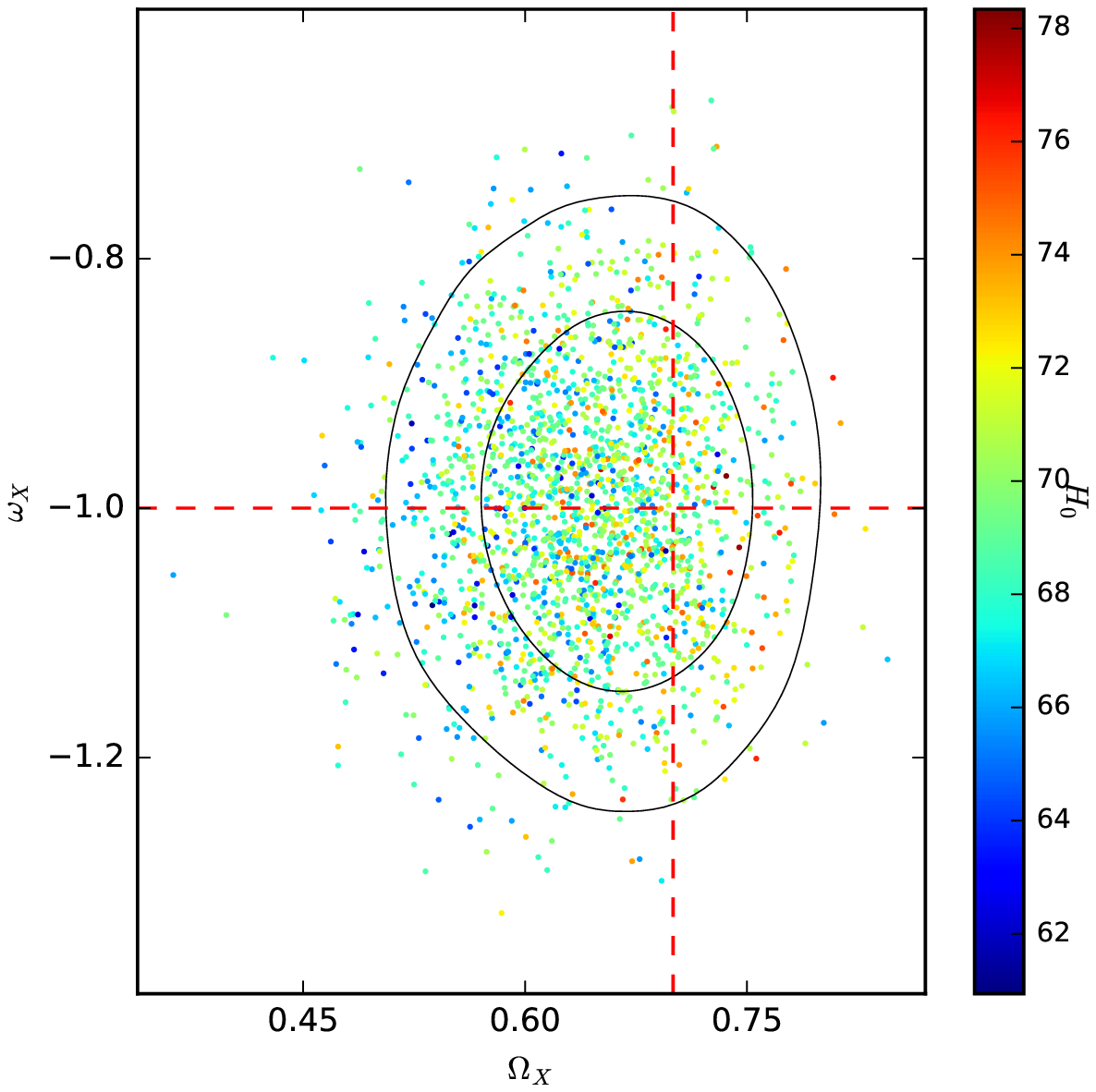}
\caption{$\omega_{X}-\Omega_{X}$ plane with $1\sigma$ and $2\sigma$ confident level in the XBI model considering two normal $H_{0}$ priors. Samples are colored by the parameter $H_{0}$. Solid contours stand for prior $H_{0}=73\pm1.74$. Red-dashed horizontal \& vertical lines stand for $\omega_{X}=-1$ \& $\Omega_{X}=0.7$.}
\label{fig9}
\end{figure}

\begin{figure}[h!]
\includegraphics[width=8.5cm,height=6cm,angle=0]{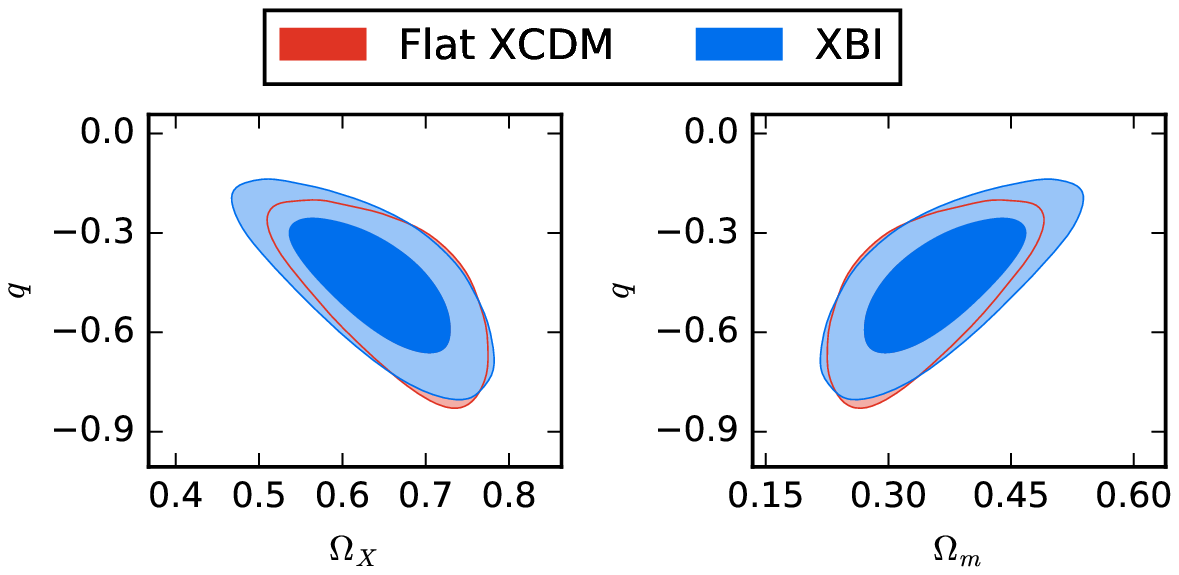} \\
\caption{$q-\Omega_{X}$ and $q-\Omega_{m}$ planes with $1\sigma$ and $2\sigma$ confident level for flat XCDM \& XBI models considering normal prior for Hubble constant as $H_{0}=68\pm2.8$.}
\label{fig10}
\end{figure}

\begin{figure}[h!]
\includegraphics[width=8.5cm,height=6cm,angle=0]{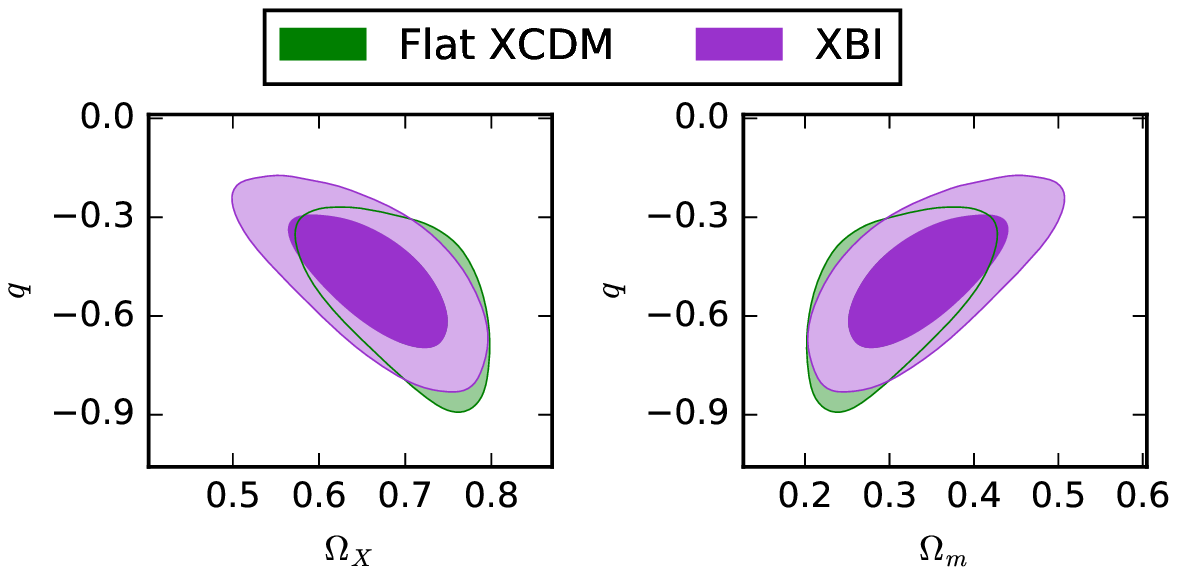}
\caption{$q-\Omega_{X}$ and $q-\Omega_{m}$ planes with $1\sigma$ and $2\sigma$ confident level for flat XCDM \& XBI models considering normal prior for Hubble constant as $H_{0}=73\pm1.74$.}
\label{fig11}
\end{figure}

The XBI model has five unknown parameter to be estimated from 36 OHD parameter measurements. The base parameters set for this model is
\begin{equation}
\label{eq24} {\bf P}=\{\Omega_{0m},\Omega_{0\tilde{K}},\omega_{X},H_{0},q\}.
\end{equation}
This model is of our most interest as it describes a flat universe such as flat XCDM model but allows a small anisotropy, $\Omega_{0\tilde{K}}$ (see \cite{ref42} for derivation of anisotropy parameter). We have shown our results of statistical analysis for XBI model in Table~\ref{tab:5}. The robustness of our fit could be viewed by looking at Fig.~\ref{fig7} which depicts the contour plots of the model parameters at $1\sigma$ and $2\sigma$ confidence regions. Again we observe that if $H_{0}=73\pm1.74$ is taken to be the prior for the current value of Hubble constant, one could get better estimations with respect to $H_{0}=68\pm2.8$ comparing to the 9 year WMAP \cite{ref10}\& Planck 2015 collaboration \cite{ref8}. The estimated valve of the anisotropy parameter for both $H_{0}$ priors is very small which is in consistence with almost all observations. From Table~\ref{tab:5} and Figs.~\ref{fig8} \& \ref{fig9} we see that $\Omega_{m}$ \& $\Omega_{X}$ are in better agreement with Planck 2016 collaboration. Depending on the $H_{0}=73\pm1.74 (68\pm2.8)$, we current value of the deceleration parameter is computed as $q=-0.48\pm 0.14 (-0.45^{+0.14}_{-0.13})$ which indicates that for first $H_{0}$ prior, XBI model behaves close to the standard cosmological model. Here, it is worth mentioning that the computed value of $\Omega_{0\tilde{K}}$ is of order $\sim 10^{-3}$ which is 100 times larger than the $\sim 10^{-5}$ level anisotropies in the CMB. This means that using OHD alone is not enough to constrain this parameter and a joint combination of other datasets, specially CMB, is required for better estimation. Figs.~\ref{fig10} \& \ref{fig11} compare the pairs $(q, \Omega_{m,X})$ of flat XCDM and XBI models considering two $H_{0}=68\pm2.8, 73\pm 1.74$ priors respectively. From these Figs it is clear that for both models choosing $H_{0}=73\pm1.74 $ puts titter restrictions on the models parameters. However, XBI model fits to the OHD dataset slightly better than flat XCDM model.

\begin{table}[h!]
\caption{Results from the fits of the XBI model to the data at 1$\sigma$, 2$\sigma$, 3$\sigma$ confidence levels.}
\centering
\setlength{\tabcolsep}{2pt}
\scalebox{0.7}{
\begin{tabular}{c p{1cm} cccc}
\hline
\hline
& & Parameter & $\%68$ & $\%95$ & $\%99$ \\
\hline
\multirow{16}{*}{$H_{0}=68\pm 2.8$} &
\multirow{1}{*}{Fit} &
$H_{0}$ &  $69.3\pm 2.7$ & $69.3^{+5.2}_{-5.2}$ &  $69.3^{+6.9}_{-6.8}$ \\[.5cm]
&  & $\Omega_{m}$ &  $0.364^{+0.059}_{-0.070}$ & $0.36^{+0.14}_{-0.12}$ &  $0.36^{+0.18}_{-0.15}$ \\[.5cm]
& & $q$ & $-0.45^{+0.14}_{-0.13}$  &  $-0.45^{+0.26}_{-0.28}$  & $-0.45^{+0.33}_{-0.37}$ \\[.5cm] 
& & $\omega_{X}$ & $-0.99\pm 0.10$ & $-0.99^{+0.20}_{-0.20}$ & $-0.99^{+0.26}_{-0.26}$ \\[.5cm]
& & $\Omega_{\tilde{K}}$ & $-0.0020\pm 0.0019$ & $-0.0020^{+0.0037}_{-0.0038}$ & $-0.0020^{+0.0052}_{-0.0051}$ \\[.5cm]
\cline{2-6}
& \multirow{2}{*}{Derived} &
$t_{0}$ & $9.62\pm 1.83$ & - & -  \\[.5cm]
& & $\Omega_{X}$ & $0.638^{+0.069}_{-0.058}$ & - & -  \\[.5cm]
\hline
\hline
\multirow{16}{*}{$H_{0}=73\pm 1.74$} &
\multirow{1}{*}{Fit} &
$H_{0}$ &  $73.3\pm 1.7$ &    $73.3^{+3.3}_{-3.3}$ &  $73.3^{+4.5}_{-4.4}$ \\[.5cm]
& &$\Omega_{m}$ &  $0.342^{+0.056}_{-0.067}$ & $0.34^{+0.13}_{-0.11}$ & $0.34^{+0.17}_{-0.14}$ \\[.5cm]
& &$q$ & $-0.48\pm 0.14$ & $-0.48^{+0.26}_{-0.27}$ & $-0.48^{+0.33}_{-0.37}$ \\[.5cm]
& &$\omega_{X}$ & $-0.99\pm 0.10 $ & $-0.99^{+0.20}_{-0.20}$ & $-0.99^{+0.26}_{-0.26}$ \\[.5cm]
& &$\Omega_{\tilde{K}}$ & $-0.0024\pm 0.0018$ & $-0.0024^{+0.0034}_{-0.0036}$ & $-0.0024^{+0.0045}_{-0.0048}$ \\[.5cm]
\cline{2-6}
& \multirow{2}{*}{Derived} &
$t_{0}$ & $8.96\pm 2.23$  &  -  & - \\[.5cm] 
& &$\Omega_{X}$ & $0.661^{+0.065}_{-0.055}$ & - & - \\[.5cm]
\hline
\hline
\end{tabular}}
\label{tab:5}
\end{table}

To compare the considered cosmologies, in what follows, we perform Akaike information criterion (AIC) and Bayes factor ($\Psi$). For a given  dataset and a given theoretical model, the Akaike estimate of minimum information (Akaike 1974) \cite{ref63} is defined as
\begin{equation}
\label{eq25} AIC=-2Log\mathcal{L}^{max}+2N,
\end{equation}
where $N$ is the number of independent parameters of the Model. Also the Bayes factor for a given dataset ($D$) and Models $M_{0}$(the null hypothesis) and $M_{1}$
\begin{equation}
\label{eq26} \Psi=\dfrac{P(D|M_{1})}{P(D|M_{0})},
\end{equation}
where $P(D|M_{1})$ is the likelihood of data given $M_{1}$ and $P(D|M_{1})$ is the likelihood of data given $M_{0}$. While in the Akaike method the preference is given to the model with the lowest AIC, Bayes method provides a criterion for choosing between two models by comparing their best likelihood values. Note that $0\leq (\Delta AIC)\leq2$ gives strong evidence for the model whereas $\Delta AIC\leq7$ indicates moderate support for the model (values higher than 10 show no evidence). Also Bayes factor ($\Psi$) represents
the odds for the null hypothesis (in our study Flat XCDM model) against other models. It is worth to mention that odds less than 1:10 show a strong evidence against XCDM model whereas odds greater than 10 : 1 indicate a strong evidence against non-flat XCDM and $\omega$BI model (Jeffreys \cite{ref64}). We have shown the difference, $\Delta AIC= AIC_{\mbox{XBI} (\mbox{AIC non-flat XCDM})}-AIC_{flat XCDM}$ and the Bayes factor $\Psi$ of both other models against flat XCDM model in Table~\ref{tab:6}. From his table we observe that when we consider $H_{0}=68\pm2.8$ as prior, XBI model fits to the data better than non-flat XCDM model.
\begin{table}[h!]
\caption{Comparison of the cosmological models by $\Delta(AIC)$ and $\Psi$ using joint OHD data}
\centering
\setlength{\tabcolsep}{2pt}
\scalebox{0.7}{
\begin{tabular}{c p{1cm} cccc}
\hline
\hline
& & Model & $\Delta(AIC)$ & $\Psi$\\
\hline
\multirow{4}{*}{$H_{0}=68\pm 2.8$} &
\multirow{1}{*}{} &
flat XCDM &  $0$ & $1$\\[.2cm]
& & non-flat XCDM & 7.4  &1:1.02\\[.2cm] 
& & XBI & 5.14 &1:1.01\\[.2cm]
\hline
\multirow{4}{*}{$H_{0}=73\pm 1.74$} &
\multirow{1}{*}{} &
flat XCDM &  0 & 1\\[.2cm]
& & non-flat XCDM & 8.4  & 1:1.02\\[.2cm] 
& & XBI & 9.06 & 1:102\\[.2cm]
\hline
\hline
\end{tabular}}
\label{tab:6}
\end{table}
\section{Cosmological Transition Redshift}\label{sec:4}
\begin{figure}[h!]
\includegraphics[width=8cm,height=6cm,angle=0]{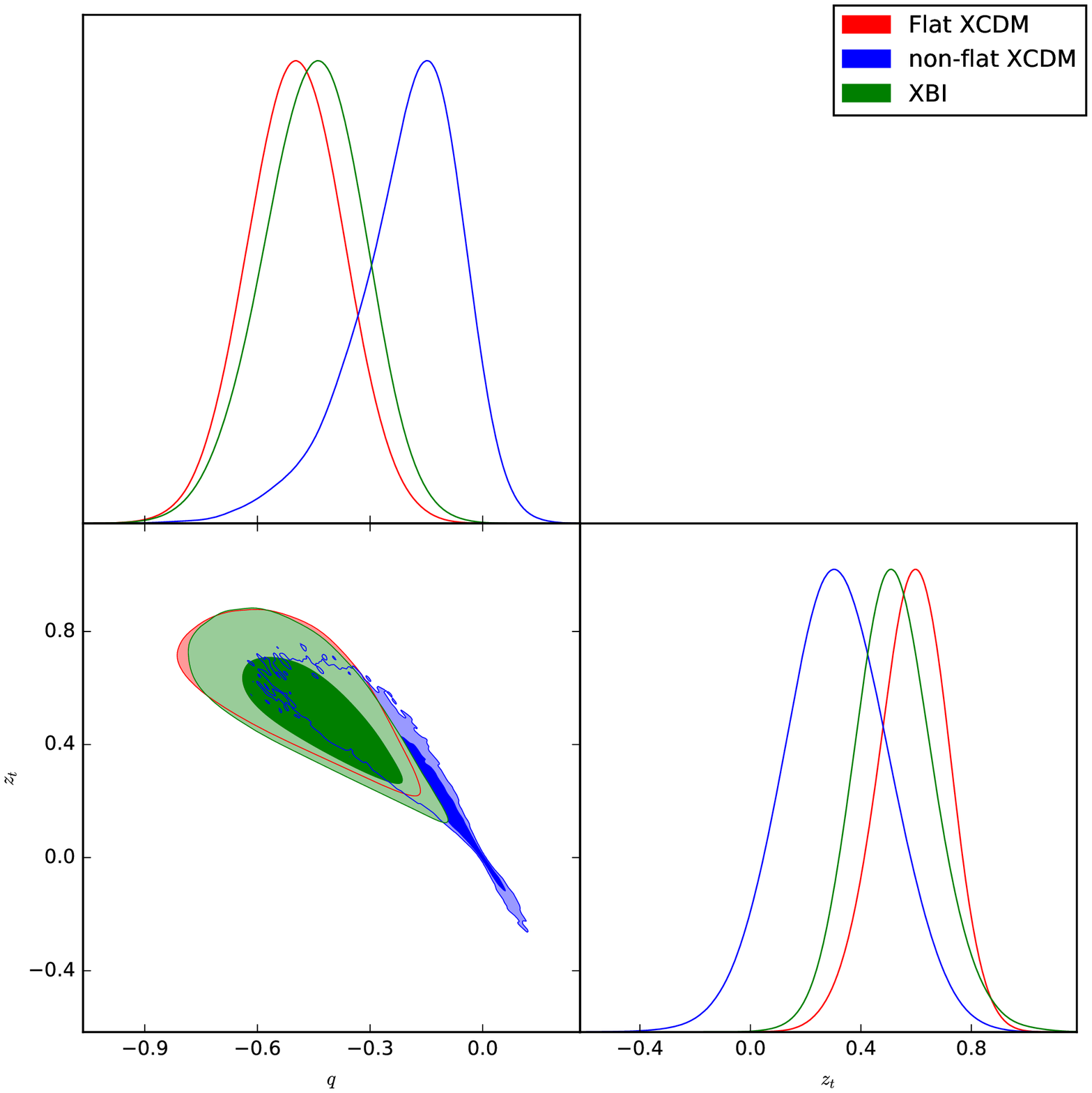} \\
\caption{One-dimensional marginalized distribution, and two-dimensional contours in $z_{t}-q$ plane with $68\%$ CL and $95\%$ CL for Flat and curved XCDM and XBI models considering considering normal prior for Hubble constant as $H_{0}=68\pm2.8$.}
\label{fig12}
\end{figure}

\begin{figure}[h!]
\includegraphics[width=8cm,height=6cm,angle=0]{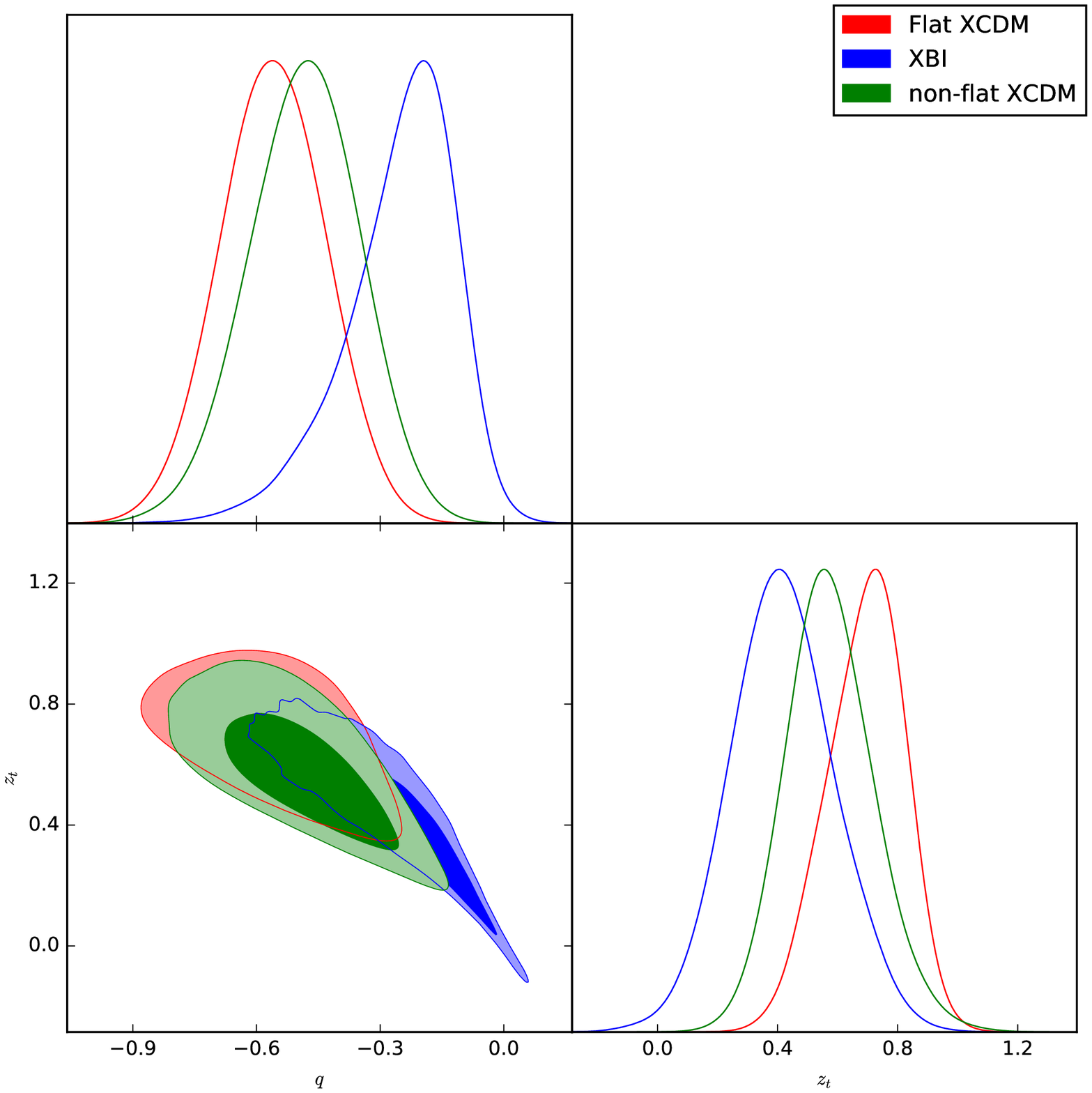}
\caption{One-dimensional marginalized distribution, and two-dimensional contours in $z_{t}-q$ plane with $68\%$ CL and $95\%$ CL for Flat and curved XCDM and XBI models considering considering normal prior for Hubble constant as $H_{0}=73\pm1.4$.}
\label{fig13}
\end{figure}

At earlier times when universe was dominated by non-relativistic matter the
cosmological expansion was slowing down. Later on, at a certain time (redshift) dark energy dominated the cosmic energy budget which in turn accelerated the cosmological expansion. The transition redshift, $z_{t}$ , is implicitly defined by the condition $q(z_{t})=\ddot{a_{t}}=0$. In this section we derive the deceleration-acceleration redshift $z_{t}$ for three DE models under consideration. In fact, this is the redshift at which the expansion phase of the universe changes from decelerating to accelerating. In general, the deceleration parameter is defined as
\begin{equation}
\label{eq27} q(z)=-\frac{1}{H^{2}}\left(\frac{\ddot{a}}{a}\right)=\frac{(1+z)}{H(z)}\frac{dH(z)}{dz}-1.
\end{equation}
Using this equation we could find a general equation of transition redshift for above mentioned three DE models as
\begin{equation}
\label{eq28} z_{t}= \left[\frac{\Omega_{0m}}{(\Omega_{0m}+\tilde{\Omega}-1)(1+3\omega_{X})}\right]^{\frac{1}{3\omega_{X}}}-1,
\end{equation}
where
\begin{equation}
\label{eq29} \tilde{\Omega}= \begin{cases}
\Omega_{0K} &\mbox{ for non- flat XCDM} \\
0 &\mbox{ for flat XCDM}\\
\Omega_{0\tilde{K}} &\mbox{for $\omega$ BI}\\
\end{cases}
\end{equation}
We have shown the deceleration-acceleration transition redshift $z_{t}$ derived from fitting of the models to the 36 $H(z)$ observational Hubble data in Table~\ref{tab:7}. From this Table we observe that for $H_{0}=73\pm1.74$ the transition from decelerating to accelerating expansion occurs at higher redshifts in all three models. For both $H_{0}$ priors we found $z_{t}(flat XCDM)>z_{t}(XBI)>z_{t}(non-flat XCDM)$. Moreover, the transition redshift for both flat XCDM and XBI models is comparable with the transition redshift of flat $\Lambda$CDM model. Figs.~\ref{fig12} \& \ref{fig13} depict one-dimensional marginalized distribution and two-dimensional contours in $z_{t}-q$ plane with $68\%$ CL and $95\%$ CL for Flat, curved XCDM, and XBI models for two Hubble constant priors $H_{0}=68\pm2.8$ and $H_{0}=73\pm1.74$ respectively. Using $36H(z)$ datapoints, Yu et al \cite{ref32} showed that the transition redshift should be restricted as $0.33<z_{t}<1$ at $1\sigma$ significance. Also using $38H(z)$ datapoints Farooq et al \cite{ref31} obtained $z_{t} = 0.72\pm0.05 (0.84\pm0.03)$ for $H_{0}= 68 \pm 2.8 (73.24 \pm 1.74) kms^{-1}Mpc^{-1}$ (Also see \cite{ref65} for the results obtained from $28H(z)$ data). Comparing our obtained transition redshift with theses results shows an excellent agreement for XCDM and XBI models.
\begin{table}[h!]
\caption{The deceleration-acceleration transition redshift $z_{t}$ at 1$\sigma$, 2$\sigma$, 3$\sigma$ CL for three models under study.}
\centering
\setlength{\tabcolsep}{2pt}
\scalebox{0.7}{
\begin{tabular}{c p{1cm} cccc}
\hline
\hline
& & Model & $\%68$ & $\%95$ & $\%99$ \\
\hline
\multirow{4}{*}{$H_{0}=68\pm 2.8$} &
\multirow{1}{*}{} &
flat XCDM &  $0.59^{+0.14}_{-0.12}$ & $0.59^{+0.24}_{-0.27}$ &  $0.59^{+0.31}_{-0.36}$ \\[.2cm]
& & non-flat XCDM & $0.31\pm0.19$  &  $0.31^{+0.37}_{-0.38}$  & $0.31^{+0.49}_{-0.51}$ \\[.2cm] 
& & XBI & $0.52^{+0.13}_{-0.15}$ & $0.52^{+0.31}_{-0.28}$ & $0.52^{+0.43}_{-0.37}$ \\[.2cm]
\hline
\multirow{4}{*}{$H_{0}=73\pm 1.74$} &
\multirow{1}{*}{} &
flat XCDM &  $0.69^{+0.15}_{-0.11}$ & $0.69^{+0.23}_{-0.26}$ &  $0.69^{+0.30}_{-0.35}$ \\[.2cm]
& & non-flat XCDM & $0.42\pm0.17$  &  $0.42^{+0.35}_{-0.33}$  & $0.42^{+0.45}_{-0.46}$ \\[.2cm] 
& & XBI & $0.57^{+0.13}_{-0.16}$ & $0.57^{+0.31}_{-0.28}$ & $0.7^{+0.43}_{-0.36}$ \\[.2cm]
\hline
\hline
\end{tabular}}
\label{tab:7}
\end{table}
\section{Conclusions}\label{sec:5}
Using recently compiled 36 observational Hubble data (OHD), in this paper, we studied and compared three dark energy models namely flat and curved FRW and Bianchi type I models. To consider the effect of the special choose of the Hubble constant value, we assumed two Gaussian priors as $H_{0}=73\pm1.74 (68\pm2.8)$. Our statistical analysis show that, in general, curved XCDM model is not in agreement with the 9years WMAP as well as Planck 2016 collaboration as far as we constrain this model to the OHD. However, we found that flat XCDM and XBI models are in good agreement with above mention observational results when fit to the OHD dataset. We also found that the transition redshift from decelerating to accelerating expansion in XBI model is in slightly better agreement with $\Lambda$CDM model with respect to flat XCDM model. To study our universe, May by, we can argue that in general XBI model is better than flat XCDM model as it allow us to study the CMB anisotropy in a natural way.
\begin{acknowledgments}
Authors are grateful to Professor Bharat Ratra for critical review of the manuscript prior to submission.
\end{acknowledgments}
						

\end{document}